# Effect of distance on photoluminescence quenching and proximity-induced spin-orbit coupling in graphene-WSe$_2$ heterostructures

Bowen Yang[1], Everardo Molina[1], Jeongwoo Kim[2], David Barroso[3], Mark Lohmann[1], Yawen Liu[1], Yadong Xu[1], Ruqian Wu[2], Ludwig Bartels[3], Kenji Watanabe[4], Takashi Taniguchi[4] and Jing Shi[1*]

[1]Department of Physics and Astronomy, University of California, Riverside, CA 92521

[2]Department of Physics and Astronomy, University of California, Irvine, CA 92697 [3]Department of Chemistry and Materials Science & Engineering Program, University of California, Riverside, CA 92521

[4]National Institute for Materials Science, 1-1 Namiki, Tsukuba, 305-0044 Japan

Corresponding author's email: jing.shi@ucr.edu; phone: (951) 827-1059.

# Abstract

Spin-orbit coupling (SOC) in graphene can be greatly enhanced by proximity coupling it to transition metal dichalcogenides (TMDs) such as $WSe_2$. We find that the strength of the acquired SOC in graphene depends on the stacking order of the heterostructures when using hexagonal boron nitride ($h$-BN) as the capping layer, i.e., $SiO_2$/graphene/$WSe_2$/$h$-BN exhibiting stronger SOC than $SiO_2$/$WSe_2$/graphene/$h$-BN. We utilize photoluminescence (PL) as an indicator to characterize the interaction between graphene and monolayer $WSe_2$ grown by chemical vapor deposition. We observe much stronger PL quenching in the $SiO_2$/graphene/$WSe_2$/$h$-BN stack than in the $SiO_2$/$WSe_2$/graphene/$h$-BN stack, and correspondingly a much larger weak antilocalization (WAL) effect or stronger induced SOC in the former than in the latter. We attribute these two effects to the interlayer distance between graphene and $WSe_2$, which depends on whether graphene is in immediate contact with $h$-BN. Our observations and hypothesis are further supported by first-principles calculations which reveal a clear difference in the interlayer distance between graphene and $WSe_2$ in these two stacks.

Transition metal dichalcogenides (TMDs) such as MoS$_2$ have attracted a great deal of attention in the two-dimensional (2D) materials community. Due to the presence of a direct band gap, monolayer TMDs are uniquely suited for optical exploration of the valley degree of freedom including the optical selection rules[1,2] and valley Hall effect (VHE)[3,4]. In the meantime, the research of graphene, an older member of the 2D material family, has achieved significant breakthroughs since its discovery[5]. For example, graphene grown by chemical vapor deposition (CVD) has been shown to have mobility[6] up to $3 \times 10^6 cm^2 s^{-1} V^{-1}$ and spin life time up to 12 ns[7].

Recently, TMD/graphene heterostructures that take advantage of unique properties of both have gained considerable research interest. Unlike conventional thin film heterostructures, TMD/graphene is formed with atomically thin layers stacking on each other via the van der Waals (vdW) interaction. The atomic flatness of these thin layers promotes strong proximity effects that modify material properties or give rise to novel interfacial phenomena. The semi-metallic nature of graphene reduces or eliminates the notorious Schottky barrier between direct TMD/metal contacts, which is desired for probing the electrical properties of TMD[8]. Conversely, the strong spin-orbit coupling (SOC) and associated spin-valley coupling in TMDs allow manipulation of spin degree of freedom in graphene. Indeed, a series of recent studies has shown that TMD can introduce strong SOC into graphene[9-13]. For example, MoS$_2$/graphene can act as a logic spin valve which can be switched on or off by tuning the Fermi level positions[14,15]. Spin polarization in graphene can be generated by optically pumping the neighboring MoS$_2$[16,17].

Although most of those works focus on the effects arising from the TMD-graphene interface, a third layer, e.g., the hexagonal boron nitride (*h*-BN), is often included in the heterostructures for various purposes. For example, when the *h*-BN is adjacent to graphene, it does not introduce much Coulomb scattering[18] as SiO$_2$ does so that the mobility of graphene is greatly enhanced[19]. As a capping layer, *h*-BN prevents TMDs such as WSe$_2$ from degrading in ambient conditions. In addition, *h*-BN functions as a highly efficient dielectric medium, which is proven critical to dual-gating bilayer graphene to demonstrate the VHE[20], and is important to a newly proposed bilayer graphene based spin valve[21]. In spite of many obvious benefits that *h*-BN brings to TMD/graphene heterostructures, the relatively strong vdW interaction between *h*-BN and graphene can adversely affect the interaction between TMD and graphene. For example, a recent work[13] reported that the stacking order of the layers is important, as weak antilocalization (WAL) can be only observed in $SiO_2$/graphene/$WSe_2$ but not in $SiO_2$/$WSe_2$/graphene/$h$-BN. The absence of WAL in the latter structure was attributed to the quasi-ballistic transport since WAL only occurs in the diffusive regime. However, this assumption fails to explain why WAL appears in a $SiO_2$/$WSe_2$/graphene device with similar mobility and even smaller size[10]. The discrepancy suggests that the presence of the $h$-BN layer may affect the interaction between TMD and graphene.

To explore acquired SOC in graphene devices, characterizing the interlayer interaction between TMD and graphene is clearly very important, especially before the full device nanofabrication. In a recent work, we demonstrated that the photoluminescence (PL) produced by monolayer TMD is quenched due to strong interaction between TMD and graphene. Here, we utilize the PL quenching as an indicator of the interaction strength to study the effect of the $h$-BN. We find that the presence of the $h$-BN layer can effectively pull graphene away from the TMD layer and consequently cause reduced SOC.

PL in monolayer TMD is much stronger than in multilayer TMD due to the direct band gap in the former[22]. However, when monolayer TMD is in intimate contact with graphene, the PL response is nearly quenched[23-25], as illustrated in Figure 1a. Due to the charge transfer between TMD and graphene, if the coupling is strong, the excited electron-hole pairs in TMD quickly recombine through the non-radiative channel due to graphene's semi-metallic Dirac bands[25]. As a result, the radiative recombination of the electron-hole pairs that produces the PL is greatly suppressed. On the contrary, when the TMD is not in intimate contact with graphene, as illustrated in Figure 1b, the charge transfer is significantly blocked owing to the much-increased interlayer distance between TMD and graphene which results in a much reduced tunneling probability. In other words, graphene just acts as an independent transparent layer and the PL in TMD is largely unaffected. Therefore, PL quenching can be conveniently used as an indicator of the interlayer distance between TMD and graphene.

Figure 1d shows the PL mapping of a $SiO_2$/graphene/$WSe_2$/$h$-BN stack. The $WSe_2$ flake is picked up by $h$-BN from a continuous monolayer $WSe_2$ sheet grown by chemical vapor deposition (CVD)[12] and transferred onto a large monolayer graphene flake that is exfoliated and placed on a $SiO_2$ substrate. The $h$-BN flake (in blue) is left on the stack after the transfer is completed. The randomly scattered yellowish speckles in Figure 1c are bubbles formed between graphene and $WSe_2$ at the interface. Strong PL is observed in the region without graphene, i.e., the SiO$_2$/WSe$_2$/h-BN region above the dotted line in Figure 1d. In contrast, in the SiO$_2$/graphene/WSe$_2$/h-BN region below the dotted line, the PL is nearly quenched except in the areas with bubbles where graphene is locally detached from WSe$_2$.

To show a quantitative comparison between the PL data from the bubbled and flat regions in the same SiO$_2$/graphene/WSe$_2$/h-BN heterostructure, we display both in Figure 1e. The solid curve is the PL spectrum averaged over the bubbled region shown in red in Figure 1f, and the dashed curve is the PL spectrum averaged over the flat region shown in blue in Figure 1f, the intensity of which differs by a factor of 40. Since the graphene directly under the bubbles is separated from WSe$_2$ by roughly ~10 nm[12], the strong PL is similar to that in SiO$_2$/WSe$_2$/h-BN. In contrast, the PL from the flat region is greatly suppressed. This stark contrast reveals that it is the intimate contact in the flat region of the SiO$_2$/graphene/WSe$_2$/h-BN stack that leads to the strong PL quenching effect. Therefore, the PL quenching in monolayer TMD can serve as an indicator of strong interlayer interaction and consequently the strong induced SOC in graphene.

Next we discuss the properties of the stack with the reverse order, i.e., SiO$_2$/WSe$_2$/graphene/h-BN, in which graphene is sandwiched between WSe$_2$ and h-BN. Figure 2a shows such a sample assembled by first picking up graphene with an h-BN flake and then transferring both together onto a monolayer WSe$_2$ island grown on SiO$_2$ by CVD. Figure 2b shows the PL mapping from three regions containing graphene (below the dot-dashed line), no graphene (above the dot-dashed line), and a multilayer WSe$_2$ seed (blue).

First, the PL contrast between SiO$_2$/WSe$_2$/h-BN and SiO$_2$/WSe$_2$/graphene/h-BN regions, i.e., across the dot-dashed line, is much lower compared with that in Figure 1d, indicating a much smaller quenching effect due to the presence of graphene. In addition, the multilayer seeding area (blue) clearly distinguishes itself from the monolayer area, due to the direct vs. indirect band gap of WSe$_2$. The greatly reduced PL from this multilayer area serves as a low intensity reference. Clearly, the relatively low contrast between WSe$_2$/h-BN and WSe$_2$/graphene/h-BN regions is not caused by any overall reduction of the PL intensity, e.g., the opacity of the graphene and h-BN flakes. Therefore we conclude that the PL quenching effect is indeed much weaker in the WSe$_2$/graphene/h-BN stack, even though WSe$_2$ is adjacent to graphene.

The relatively low contrast between the two regions is more clearly visualized in Figure 2c-e. The same data are re-plotted with three intensity ranges: (80, 100], (40, 80] and (0, 40]. These plots clearly delineate three spatially segmented regions that are WSe$_2$/h-BN, WSe$_2$/graphene/h-BN, and multilayer WSe$_2$, respectively. By comparing these plots, we find that the PL suppression still occurs in the WSe$_2$/graphene/h-BN reverse stack, but with a much lesser degree. Figure 2f shows the PL spectrum averaged over the colored regions in Figure 2c and Figure 2d. The PL intensity from WSe$_2$/graphene/h-BN is only reduced by 20% from that in WSe$_2$/h-BN without graphene which has the maximum PL intensity. As discussed earlier about Figure 1, when graphene is below WSe$_2$, the PL is nearly quenched. The great contrast in PL intensity between the two stacks suggests a significant role that the stacking order plays in the PL emission of WSe$_2$.

A natural explanation of the much-reduced PL suppression in the WSe$_2$/graphene/h-BN is that the interlayer interaction between WSe$_2$ and graphene is much weaker than that in the graphene/WSe$_2$/h-BN stack due to an increased interlayer distance in the former. To verify this assumption, we fabricate devices from the stacks and investigate the magneto-conductance (MC) effect in WSe$_2$/graphene/h-BN, which depends on the acquired SOC in graphene and therefore should be very sensitive to the interlayer interaction[11]. We conducted the measurements under the same condition as that in our previous work in devices with the opposite stacking order[12].

Figure 3a shows the MC data taken on the device carved from the lower-left region (WSe$_2$/graphene/h-BN) of the sample shown in Figure 2a at different hole densities. The WAL signal is very small compared with the universal conductance fluctuation (UCF) signal and random noises. To enhance the WAL signal-to-noise ratio, we have performed the ensemble averaging within a small carrier density range[26] and symmetrization about the zero magnetic field. The WAL signal is clearly discernable (the narrow blue region in the middle). Moreover, the carrier density dependence of the WAL feature shows a similar trend to that found in previous studies[11,12], i.e., the height of the central peak decreases as the carrier density approaches zero, accompanied by peak broadening. However, the absolute magnitude of the WAL feature is much smaller ($\sim 0.05\ e^2/h$) compared to SiO$_2$/graphene/WSe$_2$/h-BN stacks[12] ($\sim 0.2\ e^2/h$), suggesting that the induced SOC in graphene is indeed much smaller and the interlayer interaction between graphene and WSe$_2$ is much weaker.

Representative WAL curves for different carrier densities are shown in Figure 3b. We fit the curve at the carrier density of $n \approx 4.0 \times 10^{12} cm^{-2}$ according to the established procedure[27]. From graphene conductivity, we obtain the momentum relaxation rate of $\tau_p^{-1} \approx 4\ ps^{-1}$. By fitting equation (9) in ref. 27 to our MC data, we extract the inter-valley scattering rate $\tau_i^{-1} \approx$ $3 - 4\ ps^{-1}$, the spin relaxation rate $\tau_{SO}^{-1} \approx 0.2\ ps^{-1}$, and the dephasing rate $\tau_{\varphi}^{-1} \approx 0.1\ ps^{-1}$. Since $\tau_p^{-1}$ is the largest among all other relaxation rates, the assumption for equation (9) in ref. 27 is satisfied. The mobility ($\sim 12,000\ cm^2s^{-1}V^{-1}$) and the dephasing rate of this device are approximately the same as those shown in the previous work[12], but using the same fitting procedures, the obtained spin relaxation rate is at least four times smaller at the same carrier density. The induced SOC strength in graphene calculated for the Dyakonov-Perel (DP) mechanism[28] is approximately $0.6\ meV$, at least a factor of two smaller than that in the graphene/WSe$_2$/h-BN stack[12], indicating weaker SOC strength when graphene is adjacent to h-BN.

Both the PL quenching and induced SOC suggest that the TMD-graphene interlayer interaction in the two stacks with opposite orders is very different, which may be caused by the physical distance between the two layers[11]. The relatively stronger vdW interaction between graphene and h-BN can pull graphene away from the WSe$_2$ and therefore increases their interlayer distance. Below, we examine the interlayer distance between WSe$_2$ and graphene in these two stacks by first-principles calculations.

To clarify the correlation between the stacking sequence and the interlayer distance, density functional theory calculations (DFT) are performed for three different stacks: WSe$_2$/graphene (WG), h-BN/WSe$_2$/graphene (BWG), and WSe$_2$/graphene/h-BN (WGB). Those stacks are modeled with supercells consisting of $4 \times 4$ graphene (h-BN) and $3 \times 3$ WSe$_2$ cells expanded in the lateral plane, which are chosen to minimize the lattice mismatch (0.52 %). The interlayer distance between graphene and WSe$_2$, $d$, is determined using the atomic relaxation including vdW correction, as shown in Figure 4a. $d$ is found to be very close (different by ~0.9 %) between the WG and BWG stacks. This indicates that h-BN does not affect $d$ if it is not immediately adjacent to graphene, which is consistent with our experimental observation. However, in the WGB stack, $d$ is significantly larger (by ~3.5 %). The absolute difference may seem to be small, but due to the high sensitivity of tunneling and proximity effect to interlayer distance, such a seemingly small difference can produce observable consequences in PL intensity and WAL.

Figure 4b shows calculated band structures. WG and BWG have W-shaped inverted band gaps caused by the interplay of the Rashba SOC and the valley-Zeeman coupling SOC[11]. In contrast, the low energy state of WGB at the K point is noticeably different from the former two because the commensuration between graphene and h-BN introduces a large mass gap (13.3 meV) compared to that in the former two (0.4-0.5 meV) by breaking the AB sub-lattice symmetry. Taking the mass gap into consideration, we can extract the Rashba SOC by fitting the band structures (red dashed lines in (b)) to the model Hamiltonian (black solid lines in (b))[11], and the coefficients are found to be 0.37 meV, 0.37 meV and 0.16 meV for WG, BWG and WGB, respectively. Although the calculated values are smaller than the experimentally estimated values, these results clearly capture the effect of the vdW interaction on the distance between graphene and $WSe_2$, and consequently the effect on the induced SOC when $h$-BN is placed on $WSe_2$/graphene. Relatively, the Rashba SOC is reduced by nearly a factor of two depending on the position of $h$-BN in the stack.

To further highlight the effect of the distance, we calculate the Rashba SOC strength and plot it in Figure 4c as a function of $d$ in the WG stack when graphene is separated from $WSe_2$ with different distances (red triangles). In comparison, the extracted Rashba SOC from the calculations for the fully relaxed WG and WGB stacks is also shown by blue circles which fall on the $d$-dependence curve. We find that the Rashba SOC and the interlayer distance follows an inverse relation, and hence the reduced Rashba SOC observed in the $SiO_2$/$WSe_2$/graphene/$h$-BN stack can indeed be attributed to the increased interlayer distance between graphene and $WSe_2$.

In summary, we have studied the interlayer interaction between TMDs and graphene in two different vdW heterostructure, graphene/$WSe_2$/$h$-BN and $WSe_2$/ graphene/$h$-BN, by PL mapping and MC measurement. We find strong PL quenching exists in the former stack while the PL only quenches weakly in the latter stack. We attribute this difference to the increased interlayer distance between $WSe_2$ and graphene caused by the $h$-BN in the latter stack. This is further corroborated by much-reduced WAL and extracted SOC which is supported by the first principles calculations. We show that the PL quenching can be utilized as a convenient tool to infer the magnitude of the proximity effect between graphene and monolayer TMDs.

Acknowledgements: Graphene/TMD heterostructure device nanofabrication, transport and PL measurements, and data analyses were supported by DOE BES Award No. DE-FG02-07ER46351. Construction of the pickup-transfer optical microscope and device characterization were supported by NSF-ECCS under Awards No. 1202559 NSF-ECCS and No. 1610447. Theoretical work was supported by DOE Award No: DE-FG02-05ER46237 and calculations were performed on parallel computers at NERSC.

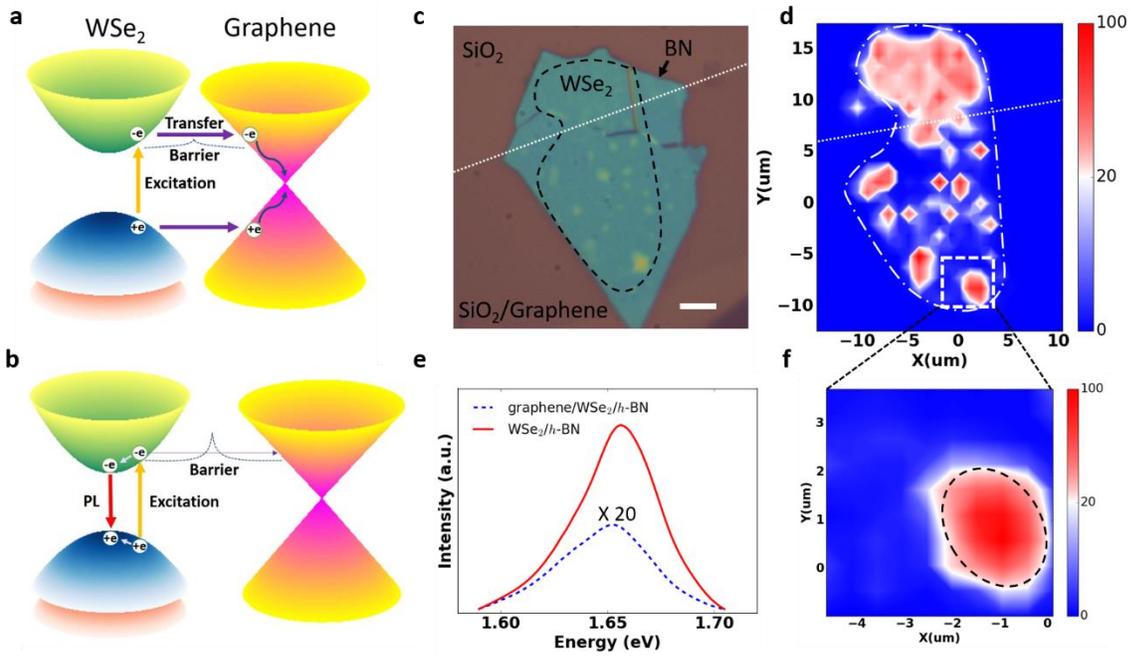

Figure 1. (a) PL quenching occurs when monolayer TMDs such as $WSe_2$ is in intimate contact with graphene and photo-excited carriers can freely move from $WSe_2$ to graphene. (b) Ordinary PL process in monolayer $WSe_2$ when graphene is far away from $WSe_2$ or absent. (c) Optical image of a $SiO_2$/graphene/$WSe_2$/$h$-BN stack. Graphene is only present below the white dotted line. $WSe_2$ picked up by $h$-BN (light blue flake) is located at the area circled by the black dashed line. Scale bar is 5 $\mu m$. (d) PL mapping of (c). The area that contains the $WSe_2$ in (c) is circled by white dot-dashed line. Dotted lines in both (c) and (d) represent the boundary separating $SiO_2$/$WSe_2$/$h$-BN (upper half) from $SiO_2$/graphene/$WSe_2$/$h$-BN (lower half). Note the PL intensity is normalized to 100. (e) Red solid curve: PL spectrum averaged over the bubbled region (area inside dashed line in (f)). Blue dashed curve: PL spectrum averaged over the flat region (area outside dashed line in (f)). Note that the dashed curve is magnified by 20 times. (f) Enlarged PL mapping of the boxed region in (d).

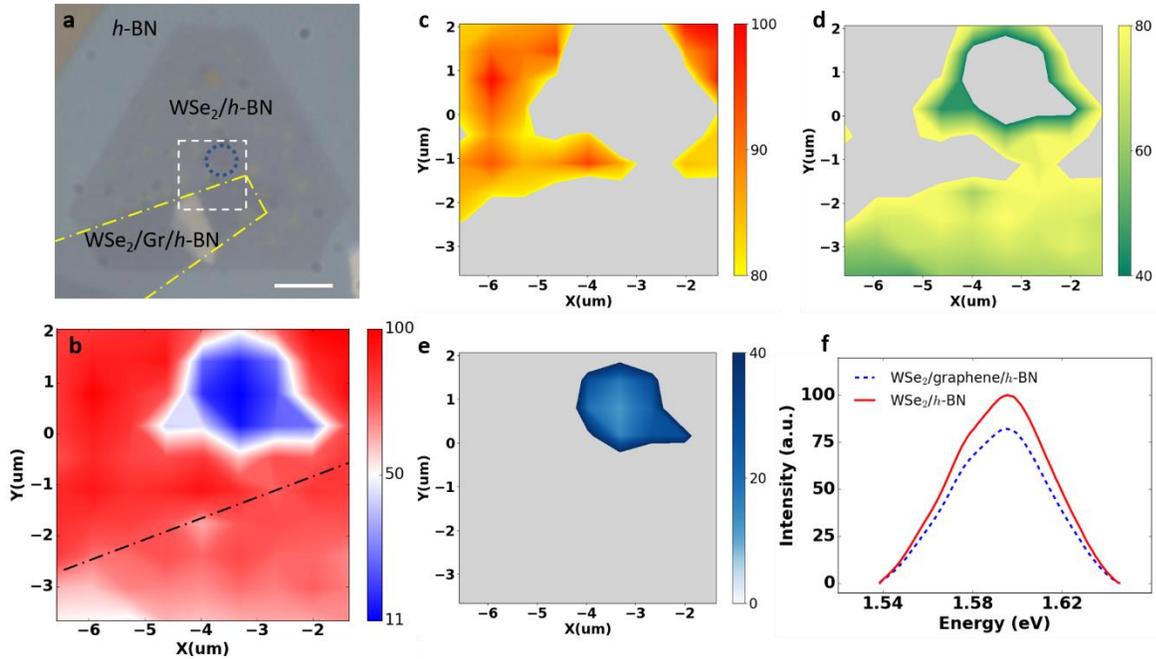

Figure 2. (a) Optical image of a SiO$_2$/WSe$_2$/graphene/h-BN stack. The h-BN flake (blueish background) occupies most of the field of view. The hexagonal-like area in the middle is the WSe$_2$ island. The graphene flake is traced by the yellow dot-dashed line. The small whitish area is a small h-BN flake on top of the large h-BN flake. A multilayer WSe$_2$ seeding area (not visible due to weak contrast but circled by the blue dotted line) lies in the middle of the monolayer WSe$_2$ island. Scale bar is 5 $\mu$m. (b) PL mapping of the area boxed in (a). The dot-dashed line separates the WSe$_2$/h-BN (dark red, upper left) and the WSe$_2$/graphene/h-BN (light red, lower right). (c-e) PL patterns plotted with the intensity falling in a particular range, i.e., (c) (80, 100] from the WSe$_2$/h-BN area; (d) (40, 80] from the WSe$_2$/graphene/h-BN area; and (e) (0, 40] from the multilayer WSe$_2$ area. Note the full scale for the original PL intensity is set to 100 and there are no data points in the grey background. (f) The PL spectrum averaged over the colored region in (c) (solid red curve) and (d) (dashed blue curve). Note that the dashed curve is not magnified.

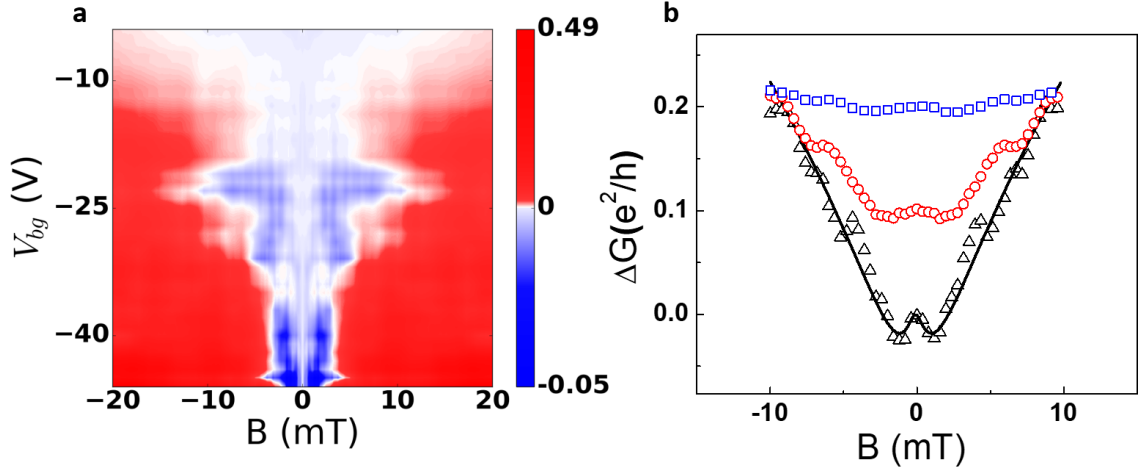

Figure 3. (a) MC measured in the WSe$_2$/graphene/$h$-BN stack on the hole side of the graphene. The blue region in the middle represents the negative MC or the WAL and the surrounding red region represents the positive MC or the weak localization (WL) Background. Note the color bar scale is different between the positive and negative values. (b) WAL curves (scattered) taken at three representative carrier densities. From bottom to up: $n \approx 4 \times 10^{12} cm^{-2}$, $3 \times 10^{12} cm^{-2}$ and $2 \times 10^{12} cm^{-2}$. The solid line is the best fitting to the WAL data. Due to the extremely small magnitude of the WAL at the other two carrier densities, we are unable to fit them reasonably.

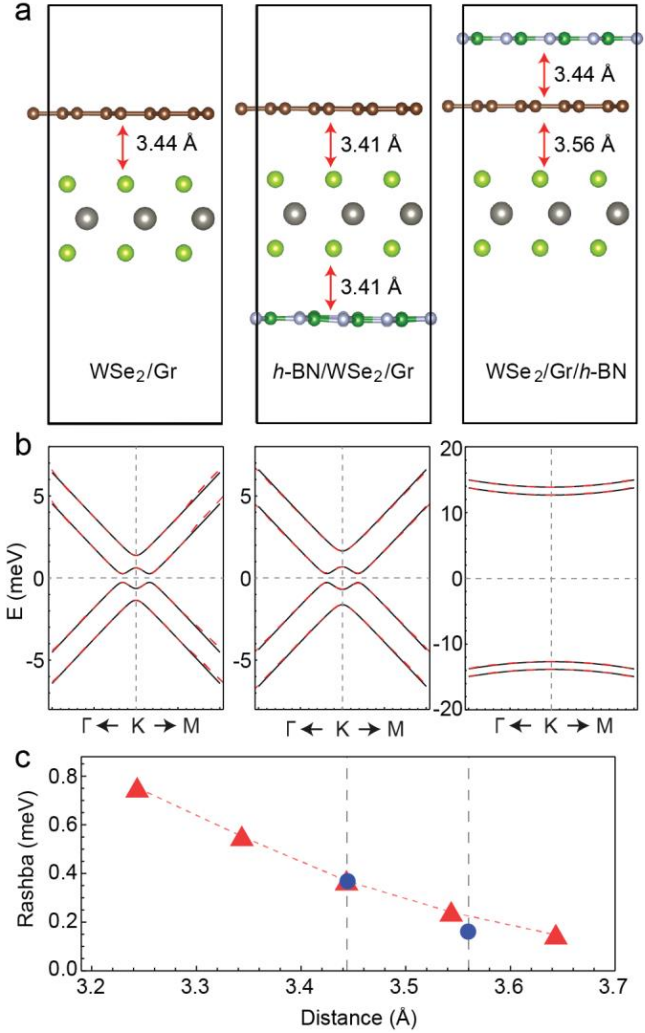

Figure 4. (a) Calculated interlayer distances between (from left to right) WSe$_2$/graphene, $h$-BN/WSe$_2$/graphene and WSe$_2$/graphene/$h$-BN stacks, respectively. (b) Calculated band structures obtained from the model Hamiltonian (black solid lines) and DFT (red dashed lines) calculations for the above stacks. (c) Red triangles: Dependence of the Rashba SOC on the interlayer distance between graphene and WSe$_2$ in the WSe$_2$/graphene stack. Blue circles: Rashba SOC extracted from relaxed WSe$_2$/graphene (left) and WSe$_2$/graphene/$h$-BN (right).

Supporting Information. Supporting information is available free of charge on the ACS Publications website at DOI: XXXX

Additional experimental details and discussion, weak and strong PL quenching data in various TMD monolayers, and vanishing weak antilocalization effect in TMD/graphene/h-BN.

TOC

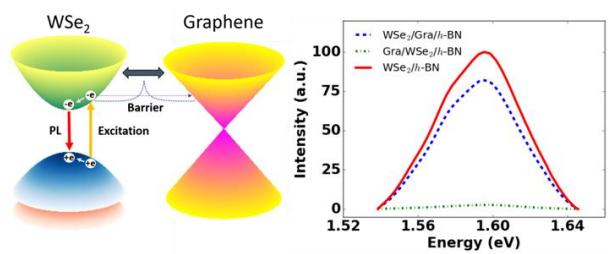